\documentclass[twocolumn]{aastex62}

\received{}
\revised{}
\accepted{}

\submitjournal{The Astrophysical Journal}

\usepackage{gensymb}

\shorttitle{Proton Acceleration in ICM Shocks}
\shortauthors{Ryu et al.}

\begin{document}

\title{A Diffusive Shock Acceleration Model for Protons in Weak Quasi-parallel Intracluster Shocks}

\author[0000-0002-5455-2957]{Dongsu Ryu}
\affiliation{Department of Physics, School of Natural Sciences, UNIST, Ulsan 44919, Korea}
\author[0000-0002-4674-5687]{Hyesung Kang}
\affiliation{Department of Earth Sciences, Pusan National University, Busan 46241, Korea}
\author[0000-0001-7670-4897]{Ji-Hoon Ha}
\affiliation{Department of Physics, School of Natural Sciences, UNIST, Ulsan 44919, Korea}
\correspondingauthor{Hyesung Kang}
\email{hskang@pusan.ac.kr}

\begin{abstract}


Low sonic Mach number shocks form in the intracluster medium (ICM) during the formation of the large-scale structure of the universe.
Nonthermal cosmic-ray (CR) protons are expected to be accelerated via diffusive shock acceleration (DSA) in those ICM shocks,
although observational evidence for the $\gamma$-ray emission of hadronic origin from galaxy clusters has yet to be established.  
{
Considering the results obtained from recent plasma simulations,
we improve the analytic test-particle DSA model for weak quasi-parallel ($Q_\parallel$) shocks, previously suggested by \citet{kang2010}.
In the model CR spectrum,
the transition from the postshock thermal to CR populations occurs at the injection momentum, $p_{\rm inj}$, above which protons can undergo the full DSA process.
As the shock energy is transferred to CR protons, the postshock gas temperature should decrease accordingly and
the subshock strength weakens due to the dynamical feed of the CR pressure to the shock structure. 
This results in the reduction of the injection fraction, although the postshock CR pressure approaches an asymptotic value when
the CR spectrum extends to the relativistic regime.
Our new DSA model self-consistently accounts for such behaviors and adopts better estimations for $p_{\rm inj}$.}
With our model DSA spectrum, the CR acceleration efficiency ranges $\eta\sim10^{-3}-0.01$ for supercritical, $Q_\parallel$-shocks with sonic Mach number $2.25\lesssim  M_{\rm s}\lesssim5$ in the ICM. Based on \citet{ha2018b}, on the other hand, we argue that proton acceleration would be negligible in subcritical shocks with $M_{\rm s}<2.25$.

\end{abstract}


\section{Introduction}
\label{s1}

Hierarchical clustering of the large-scale structure of the universe induces supersonic flow motions of baryonic matter, which result in the formation of weak shocks with sonic Mach numbers $M_{\rm s} \lesssim 4$ in the hot intracluster medium (ICM) \citep[e.g.,][]{ryu2003,vazza2009,ha2018a}.
In particular, shocks associated with mergers of subcluster clumps have been observed in X-ray and radio \citep[e.g.,][]{brunetti2014, vanweeren19}. These ICM shocks are thought to accelerate cosmic ray (CR) protons and electrons via diffusive shock acceleration (DSA) \citep{bell1978,drury1983}.
Although the acceleration of relativistic electrons can be inferred from the so-called giant radio relics \citep[e.g.,][]{vanweeren19}, the presence of the CR protons produced by ICM shocks has yet to be established \citep[e.g.,][]{pfrommer2004,pinzke2010,zandanel2014,vazza2016,kang2018}.
The inelastic collisions of CR protons with thermal protons followed by the decay of $\pi^0$ produce diffuse $\gamma$-ray emission, which has not been detected so far \citep{ackermann2016}.
Previous studies using cosmological hydrodynamic simulations with some prescriptions for CR proton acceleration
suggested that the non-detection of $\gamma$-ray emission from galaxy clusters would constrain the acceleration efficiency
$\eta \lesssim 10^{-3}$ for ICM shocks with $2\lesssim M_{\rm s}\lesssim 5$ \citep[e.g.,][]{vazza2016};
the acceleration efficiency is defined in terms of the shock kinetic energy flux, as
$\eta\equiv E_{\rm CR,2} u_2/(0.5\rho_1 u_{\rm sh}^3)$ \citep{ryu2003}.
Hereafter, the subscripts $1$ and $2$ denote the preshock and postshock states, respectively.
And $\rho$ is the density, $u$ is the flow speed in the shock-rest frame, $u_{\rm sh}$ is the shock speed, and $E_{\rm CR,2}$ is the postshock CR proton energy density.

Proton injection is one of the key processes that govern the DSA acceleration efficiency.
In the so-called thermal leakage model, suprathermal particles in the tail of the postshock thermal distribution were thought to re-cross the shock from downstream to upstream and participate in the Fermi I process \citep[e.g.][]{malkov97, kang2002}.
Through hybrid simulations, however, \citet[][CS14a, hereafter]{caprioli2014a} showed that 
in quasi-parallel ($Q_\parallel$, hereafter, with $\theta_{\rm Bn}\lesssim 45\degree$) shocks,
protons are injected through specular reflection off the shock potential barrier,
gaining energy via shock drift acceleration (SDA), and that the self-excitation of upstream turbulent waves is essential for 
multiple cycles of reflection and SDA.
Here, $\theta_{\rm Bn}$ is the obliquity angle between the shock normal and the background magnetic field direction.
They considered relatively strong ($M_{\rm s}\gtrsim 6.5$) $Q_\parallel$-shocks in plasmas with $\beta \sim 1$,
where $\beta = P_{\rm gas}/P_{\rm B}$ is the ratio of the gas to magnetic pressures.
As CRs are accelerated to higher energies, the CR energy density increases in time 
before the acceleration saturates at $E_{\rm CR,2}/(E_{\rm CR,2}+ E_{\rm th,2})\approx 0.06-0.13$ for $M_{\rm s}\approx 6.3-63$ (see Figure 3 of CS14a)
with the injection fraction, $\xi \sim 10^{-4}-10^{-3}$ (see Equation [\ref{inj}] below).
Here, $E_{\rm th,2}$ is the energy density of postshock thermal protons. 
As a result, the postshock thermal distribution gradually shifts to lower temperatures as the CR power-law tail increases its extent (see Figure 1 of CS14a).

Moreover, CS14a found that in the immediate postshock region, the proton momentum distribution can be represented by three components: the Maxwellian distribution of thermal particles,
$f_{\rm th}(p)$, the CR power-law spectrum, $f_{\rm CR}(p)$, and the suprathermal `bridge' connecting smoothly $f_{\rm th}$ and $f_{\rm CR}$ (see Figure 2 of CS14a).
This suprathermal bridge gradually disappears as the plasma moves further downstream away from the shock, because the electromagnetic turbulence
and ensuing kinetic processes responsible for the generation of suprathermal particles decrease in the downstream region.
Far downstream from the shock, the transition from the Maxwellian to CR distributions occurs 
rather sharply at the so-called injection momentum, which can be parameterized as 
$p_{\rm inj} \approx Q_{\rm i}\ p_{\rm th,p}$, where $p_{\rm th,p} = \sqrt{2m_p k_B T_2}$ is the postshock thermal proton momentum and $Q_{\rm i}\sim 3-3.5$ is the injection parameter.
Here, $T_2$ is the temperature of postshock thermal ions, $m_p$ is the proton mass, and $k_B$ is the Boltzmann constant.
They suggested that the CR energy spectrum can be modeled by the DSA power-law attached to the postshock Maxwellian at $p_{\rm inj}$,
although their hybrid simulations revealed a picture that is quite different from the thermal leakage injection model.
Later, \citet[][CPS15, hereafter]{caprioli2015} presented a minimal model for proton injection that accounts for quasi-periodic shock reformation
and multicycles of reflection/SDA energization, and predicted the CR spectrum consistent with the hybrid simulations of CS14a.

Recently, \citet[][HRKM18, hereafter]{ha2018b} studied, through Particle-in-Cell (PIC) simulations, the early acceleration of CR protons in weak ($M_{\rm s} \approx 2 - 4$) $Q_\parallel$-shocks in hot ICM plasmas where $\beta \sim 100$ \citep[e.g.,][]{ryu2008}.
In the paper, they argued that only supercritical $Q_\parallel$-shocks with $M_{\rm s}\gtrsim2.25$ develop overshoot/undershoot oscillations in their structures, resulting in a significant amount of incoming protons being reflected at the shock and injected into the DSA process. 
Subcritical $Q_\parallel$-shocks with $M_{\rm s}\lesssim2.25$, on the other hand, have relatively smooth structures,
and hence the preacceleration and injection of protons into DSA are negligible.
Thus, it was suggested that ICM $Q_\parallel$-shocks may accelerate CR protons only if $M_{\rm s}\gtrsim2.25$.
{
Although the simulations followed only to the very early stage of DSA where the maximum ion momentum reaches up to $p_{\rm max}/m_ic \sim 0.5$ ($m_i$ is the reduced ion mass\footnote{Throughout the paper, we differentiate $m_i$ from $m_p$, because the effects of the reduced mass ratio, $m_i/m_e$ ($m_e$ is the electron mass), in PIC simulations remain to be fully understood. In the simulations of HRKM18, for example, $m_i=100-800\ m_e$ was used.}),
HRKM18 attempted to quantify proton acceleration at ICM $Q_\parallel$-shocks.
The simulated CR spectrum indicated the injection parameter of $Q_{\rm i}\approx 2.7$, which led to a rather high injection fraction, $\xi \approx 2\times 10^{-3} - 10^{-2}$, for shocks with $M_{\rm s} = 2.25 - 4$.
If we simply extrapolate this injection fraction to the relativistic regime of $p_{\rm max}/m_ic\gg 1$, 
the ensuing DSA efficiency would be rather high, $\eta > 0.01$, 
which is in strong disagreement with the existing observations of $\gamma$-rays from galaxies clusters.}

{
In a `fluid-version' of numerical studies of DSA, on the other hand, the time-dependent diffusion-convection equation for the isotropic part of the momentum distribution function, $f_{\rm CR}(p)$, is solved, adopting 
a Bohm-type spatial diffusion coefficient ($\kappa\propto p$) and a `macroscopic' prescription for thermal leakage injection ($\tau_{\rm esp}$) \citep[e.g.,][]{kang2002}.
Previous studies using this approach managed to follow the
evolution of CR proton spectrum into the relativistic energies of up to $p_{\rm max}/m_pc \sim 50$ for shocks with a wide range of sonic Mach numbers \citep[e.g.,][]{kang2005}.
They showed that, as the CR pressure increases in time, the subshock weakens and $T_2$ decreases accordingly,
resulting in the gradual reduction of the injection rate and $f_{\rm CR}(p_{\rm inj})$ [see Figure 5 of \citet{kang2005}].
This leads to the decrease of the injection fraction $\xi(t)$ with time,
although the postshock CR pressure reaches an approximate time-asymptotic value [see Figure 6 of \citet{kang2002}].
These results are consistent with those of the hybrid simulations described above.}

{
Previously, \citet{kang2010} considered an analytic model for $f_{\rm CR}(p)$ in the test-particle regime of DSA for weak ICM shocks. 
They suggested that the test-particle solution of $f_{\rm CR}(p)$ could be valid only if $Q_i\gtrsim 3.8$, which results in
the injection fraction $\xi \lesssim 10^{-3}$ and the CR pressure $P_{\rm CR,2}/\rho_1 u_{\rm sh}^2 < 0.1$.
In that study, however, the changes of $T_2(t)$ and $\xi(t)$ with the increase of $p_{\rm max}$ were not included self-consistently, 
because $Q_{\rm i}$, although a free parameter, has a fixed value, and $T_2$ was estimated simply from the Rankine-Hugoniot relation, relying on the test-particle assumption. 
Hence, the model failed to incorporate the full aspect of DSA observed in the previous simulations.}

{
Based on the earlier studies of DSA using hybrid, PIC, and fluid simulations, we here propose an improved analytic model 
that is designed to approximately emulate the CR proton spectrum of DSA for given shock parameters.
The basic formulation is still based on the test-particle solution with a thermal leakage injection recipe with a free parameter, $Q_{\rm i}$,
as in \citet{kang2010}.
The main improvement is, however, the inclusion of the reduction of the postshock thermal energy density due to the transfer of the shock energy
to the CR population in a self-consistent manner; also the model considers a more realistic range of $Q_{\rm i} \approx 3.0 - 3.5$ that reflects the results of the hybrid simulations of
CS14a and CPS15.

In the next section, we first review what has been learned about proton injection and acceleration at $Q_\parallel$-shocks from recent plasma simulations.}
In Section \ref{s3}, we describe our analytic DSA model for the CR proton spectrum produced at weak $Q_\parallel$-shocks,
along with the injection fraction and acceleration efficiency that characterize the DSA of CR protons.
A brief summary follows in Section \ref{s4}.

\begin{figure*}[t]
\vskip -0.4 cm
\hskip -0.6 cm
\centerline{\includegraphics[width=1.2\textwidth]{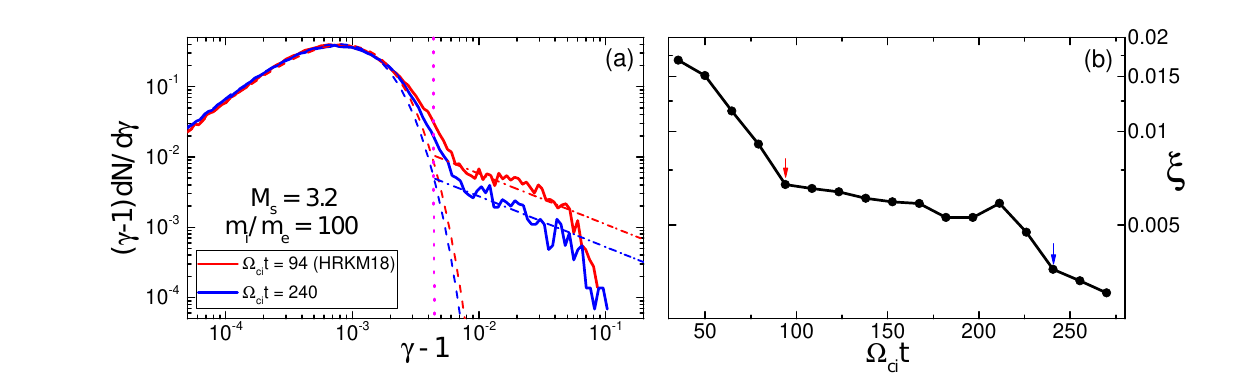}}
\vskip -0.1 cm
\caption{\label{f1}
(a) Postshock energy spectrum, $dN/d\gamma$, of ions with $m_i=100m_e$, taken from PIC simulations 
for the ICM shock of $M_s=3.2$ with $\theta_{\rm Bn}=13^{\circ}$, $\beta=100$, and $T_1=8.6$ keV ($10^8$ K). 
For $\Omega_{\rm ci}t \approx 94$~(red), the simulation data reported in HRKM18 are adopted, 
while for $\Omega_{\rm ci}t \approx 240$ (blue), those from the new extended simulation described in Section \ref{s2.2} are used.
The red and blue dashed lines show the fits for the respective spectra (solid lines) to Maxwellian and test-particle power-law forms.
The vertical dotted magenta line marks the injection energy, $\gamma_{\rm inj}$, where the two fitting forms cross each other.
(b) Time evolution of the injection fraction $\xi(t)$, calculated with the postshock energy spectra for the shock model shown in panel (a). The red and blue arrows denote the points for $\Omega_{\rm ci}t \approx 94$, and 240, respectively.}
\end{figure*}

{
\section{Implications from Plasma Simulations}
\label{s2}

Although the structure and time variation of collisionless shocks are primarily governed by the dynamics of reflected protons and the waves excited by them in the foreshock region,
the roles of electron kinetic processes in proton injection to DSA has not yet been fully explored \citep[e.g.,][]{balogh13}.
Only PIC simulations can follow from first principles various microinstabilities and wave-particle interactions due to ion and electron kinetic processes.
Owing to greatly disparate time and length scales of ion and electron processes, however, the runs of PIC simulations are limited to only several $\times10^2\ \Omega_{\rm ci}^{-1}$, depending on $m_i/m_e$, $\beta$, and the dimension of simulations.
Here, $\Omega_{\rm ci}^{-1} = {m_i c}/{eB_0}$, is the ion cyclotron period where $c$ is the speed of light, $e$ is the electron charge, 
and $B_0$ is the background magnetic field strength.
Typically, the injection and early acceleration of protons can be followed up to the maximum momentum of $p_{\rm max} /p_{\rm th,i}\sim30$ ($p_{\rm th,i} = \sqrt{2m_i k_B T_2}$) in PIC simulations \cite[e.g.,][HRKM18]{park2015}.

Hybrid simulations, in which electrons are modeled as a charge-neutralizing fluid,
can be run to several $\times10^2-10^3\ \Omega_{\rm cp}^{-1}$ (where $\Omega_{\rm cp}^{-1} = {m_p c}/{eB_0}$), neglecting details of electron kinetic processes.
Yet they can follow proton acceleration only up to $p_{\rm max}/m_pu_{\rm sh}\sim30$ or so (e.g., CP14a).

With currently available computational resources, both PIC and hybrid simulations can only study the early development of suprathermal and nonthermal protons.
Thus, it would be a rather challenging task to extrapolate what we have learned about DSA from existing plasma simulations to the relativistic regime of $p_{\rm max}/m_pc\gg 1$.

\subsection{Hybrid Simulations}
\label{s2.1}

As discussed in the introduction, the injection and acceleration of protons at $\beta\approx 1$ $Q_\parallel$-shocks with $M_{\rm s}\gtrsim 6.3$
were studied extensively through 2D hybrid simulations (CS14a and CPS15).
A small fraction of incoming protons can be injected to DSA after undergoing two to three cycles of SDA, followed by reflection off the shock potential drop.
In addition, at low-$\beta$ ($\beta\lesssim1$) shocks, the proton reflection can be facilitated by the magnetic mirror force due to the compression of 
locally perpendicular magnetic fields in upstream MHD turbulence,
which are self-excited by back-streaming protons \citep[e.g.,][]{sundberg2016}. 

The efficiency of proton injection could be quantitatively different at weak ICM shocks with $\beta\sim100$,
because the shock potential drop is smaller at lower $M_{\rm s}$ shocks and the magnetic mirror force is weaker in higher $\beta$ plasmas.
\citet[][CS14b, hereafter]{caprioli2014b}, on the other hand, showed that the magnetic field amplification due to resonant and 
non-resonant streaming instabilities increases with the Alfv\'en Mach number, $M_{\rm A}\approx \beta^{1/2} M_{\rm s}$.
Hence, the level of upstream turbulence is expected to be higher for higher $\beta$ shocks at a given $M_{\rm s}$.
Therefore, higher $\beta$ could have two opposite effects on the efficiency of proton injection, i.e., weaker magnetic mirror but stronger turbulence in the foreshock.
Unfortunately, so far hybrid simulations for high-$\beta$ ($\beta\gg1$) shocks have not been published in the literature yet.

CPS15 suggested that the proton injection at weak shocks may be different from their findings for strong shocks in the following senses: 
(1) the overshoot in the shock potential is smaller at weaker shocks, leading to a smaller reflection fraction at each confrontation with the shock,
(2) the fractional energy gain at each SDA cycle is smaller, so more SDA cycles are required for injection, 
(3) the levels of turbulence and magnetic field amplification are weaker.
As a result, the proton injection and acceleration efficiencies should be smaller at weaker shocks.
According to Figure 3 of CS14a, for the $M_{\rm s}\approx 6.3$ shock ($M=5$ in their definition), the DSA efficiency is $\eta \approx 0.036$,
so a smaller $\eta$ is expected for ICM shocks with $M_{\rm s}\lesssim 4$.

Moreover, CS14b showed in their Figure 9 that the normalization (amplitude) of postshock $f_{\rm CR}$
decreases as $p_{\rm max}(t)$ increases with time. 
We interpret that this trend is caused by the increase in the number of SDA cycles required for injection to DSA, 
because the subshock weakens gradually due to the CR feedback, and so the energy gain per SDA cycle is reduced. 
Considering that the ratio of $p_{\rm max}/p_{\rm th,p}$ reaches only to $\sim 30$ in these hybrid simulations, 
the normalization of $f_{\rm CR}$ may continue to decrease as the CR spectrum extends to the relativistic region with $p_{\rm max}/m_p c\gg 1$.

\subsection{Particle-in-cell Simulations}
\label{s2.2}

HRKM18 explored for the first time the {\it criticality} of high-$\beta$ $Q_\parallel$ shocks and showed that 
protons can be injected to DSA and accelerated to become CRs only at supercritical shocks with $M_{\rm s} \gtrsim 2.25$.
Figure 7 of HRKM18 showed that the shock criticality does not sensitively depend on $m_i/m_e$ and numerical resolution,
but the acceleration rate depends slightly on $\beta$.
As mentioned before, turbulence is excited more strongly for higher $\beta$ cases due to higher $M_{\rm A}$.
But the reflection fraction is smaller for higher $\beta$ due to weaker magnetic mirror forces, leading to lower reflection fraction and 
lower amplitude of $f_{\rm CR}$ near $p_{\rm inj}$.

In order to get a glimpse of the long-term evolution of the CR proton spectrum,
we extend the 1D PIC simulation reported in HRKM18 from $\Omega_{\rm ci} t_{\rm end}= 90$ to 270
for the model of $M_{\rm s}=3.2$, $\theta_{\rm Bn}=13^{\circ}$, $m_i/m_e=100$, $\beta=100$, and $T_1=8.6$ keV ($10^8$ K). 
Details of numerical and model setups can be found in HRKM18 (see their Table 1).
The main change is that a different computation domain, $[L_x, L_y] = [3\times 10^4,1]~(c/w_{\rm pe})^2$, is adopted here
in order to accommodate the longer simulation time.
Because of severe computational requirements, in practice, it is difficult to extend this kind of PIC simulations to a much larger box for a much longer duration.
In this simulation the average velocity of ions is $\sqrt{18.36}$ times higher than that of real protons for the given temperature.

Figure \ref{f1} shows the time evolution of the postshock energy spectra of ions, $dN/d\gamma$ (where $\gamma$ is the Lorentz factor),
and the injection fraction, $\xi(t)$ [see Eq. (11) of HRKM18].
We adopt the simulation data of HRKM18 for $\Omega_{\rm ci}t \approx 94$~(red), while the data from the new extended simulation is used for 
$\Omega_{\rm ci}t \approx 240$~(blue).
The region of $(1.5 - 2.5) r_{L,i}$ behind the shock is included, where $r_{L,i}$ is the ion Larmor radius defined with the incoming flow speed.
Note that the spectrum near the energy cutoff might not be correctly reproduced due to the limited size of the simulation domain.

We notice the following features in Figure \ref{f1}(a) :
(1) the postshock temperature decreases slightly with time,
(2) the injection parameter, $Q_{\rm i}=p_{\rm inj}/p_{\rm th,i}$, increases from $\sim2.7$ to $\sim3.0$ 
as the time increases from $\Omega_{\rm ci} t \approx 90$ to 240, and
(3) the amplitude of $dN/d\gamma (\gamma_{\rm inj})$ decreases gradually. 

Figure \ref{f1}(b) shows the resulting gradual decrease of $\xi(t)$, which may continue further in time. 
As in HRKM18, a somewhat arbitrary value of $p_{\rm min}=\sqrt{2}p_{\rm inj}$ is adopted (see the next section for a further discussion).
We interpret the bump in the evolution of $\xi(t)$ near $\Omega_{\rm ci} t \approx 210$ as a consequence of shock reformation.

\begin{figure*}[t]
\vskip -0.8cm
\hskip 0 cm
\centerline{\includegraphics[width=1\textwidth]{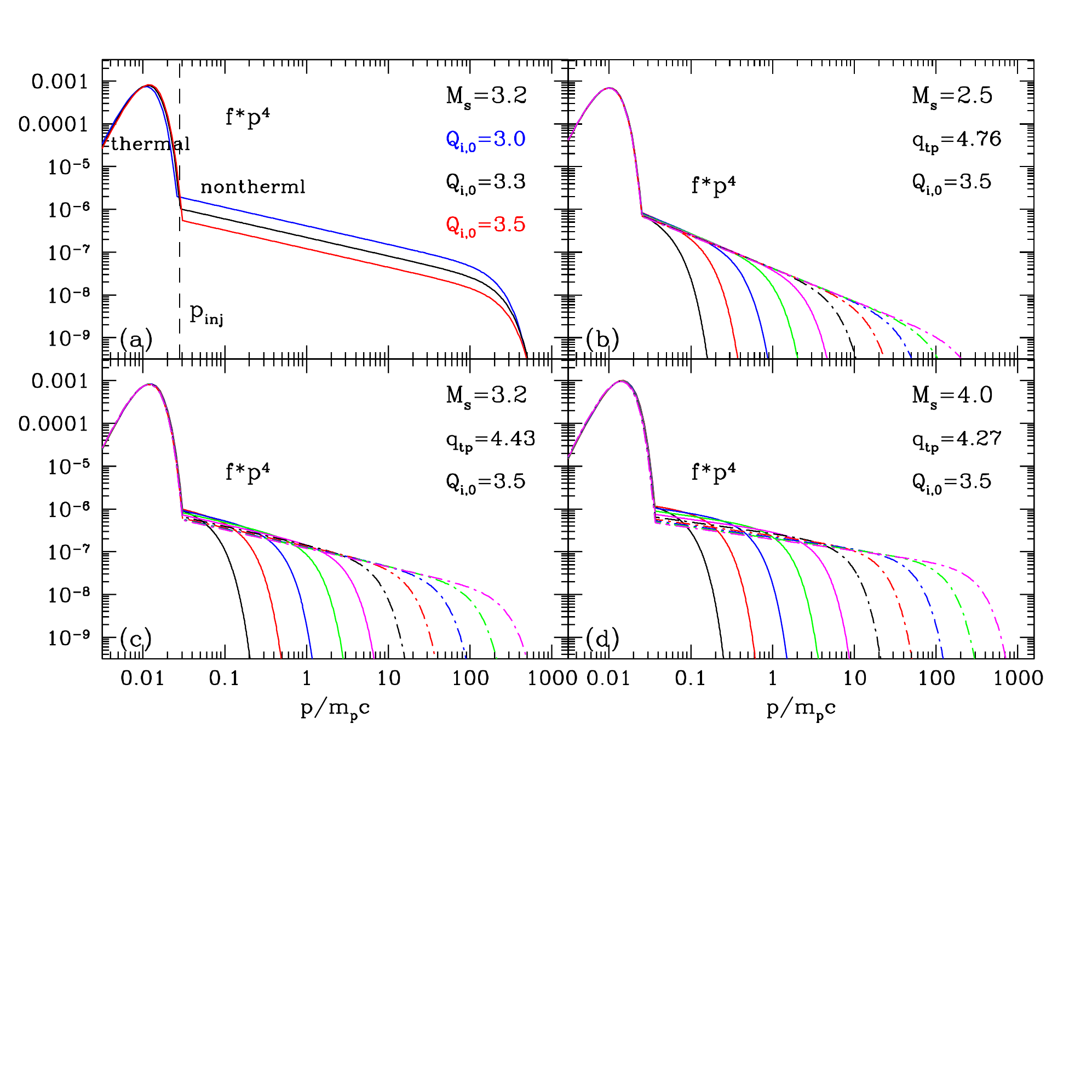}}
\vskip -6.2 cm
\caption{\label{f2} 
Proton distribution function, $f(p)p^4$, calculated with Equations (\ref{finj})-(\ref{fN}).
Panel (a): $f(p)p^4$ in a $M_{\rm s}=3.2$ shock with $Q_{\rm i,0}=$ 3.0 (blue line), 3.3 (black line), and 3.5 (red line), when the maximum momentum is $p_{\rm max} \gg p_{\rm inj}$.
The vertical dashed line shows the injection momentum, $p_{\rm inj}$, with $Q_{\rm i,0}=3.3$.
Panels (b)-(d): Change of $f(p)p^4$ in $M_{\rm s}=2.5$, 3.2, and 4.0 shocks with $Q_{\rm i,0}=3.5$, as $p_{\rm max}$ increases. Here, $T_1=10^8$ K.
{\color{red}The DSA test-particle slope, $q_{\rm tp}$, is given in each panel.}
Due to the energy transfer to the CR component, the temperature reduction factor, $R_{\rm T}$, decreases.
Hence, while $p_{\rm inj}$ is fixed, the injection parameter, $Q_{\rm i}=Q_{\rm i,0}/\sqrt{R_{\rm T}}$, increases, leading to the reduction of the normalization factor, $f_{\rm N}$. }
\end{figure*}

\section{Analytic Model for CR Proton Spectrum}
\label{s3}

\begin{figure*}[t]
\vskip -0.8 cm
\hskip -0.2 cm
\centerline{\includegraphics[width=1.05\textwidth]{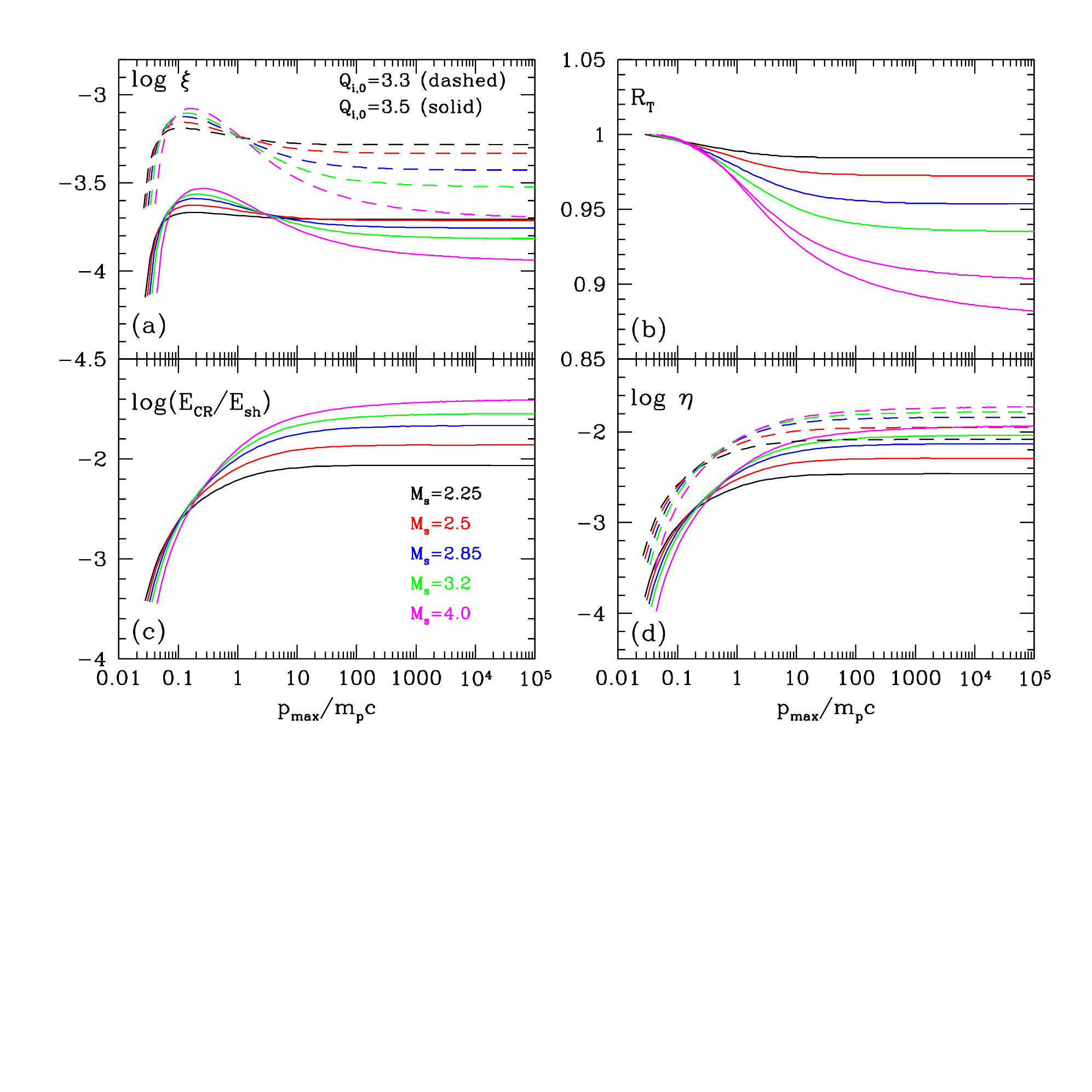}}
\vskip -6.5 cm
\caption{\label{f3}
Change of the injection fraction, $\xi$, the temperature reduction factor, $R_{\rm T}$,
the postshock CR energy fraction, $E_{\rm CR,2}/E_{\rm sh}$, and the CR acceleration efficiency, $\eta$,
as $p_{\rm max}$ increases.
Here, $T_1=10^8$ K, $p_{\rm min}=p_{\rm inj}$, $Q_{\rm i,0}=3.3$ (dashed lines) and 3.5 (solid lines) are adopted.
As $R_{\rm T}$ decreases, the injection parameter increases as $Q_{\rm i} =Q_{\rm i,0}/\sqrt{R_{\rm T}}$, 
which results in the reduction of $f_{\rm N}$ as in Equation (\ref{fN}).}
\end{figure*}

{
The analytic model presented here inherits the test-particle DSA model with a thermal leakage injection recipe, which was suggested by \citet{kang2010}. 
It describes the downstream CR proton spectrum, $f_{\rm CR}(p)$, for weak shocks.}
For the preshock gas with the density, $n_1$, and the temperature, $T_1$,
the postshock vales, $n_2$, and $T_{2,0}$\footnote{Here, $T_{2,0}$ denotes the temperature of the thermal gas when the postshock CR energy density, $E_{\rm CR,2}$, is negligible, reserving $T_2$ for the cases of non-negligible $E_{\rm CR,2}$.}, can be calculated from the Rankine–Hugoniot jump condition.
For example, the shock compression ratio is given as $r = n_2/n_1 =  (\gamma_{\rm g} + 1)/(\gamma_{\rm g} - 1 + 2/M^2_{s})$, where $\gamma_{\rm g}=5/3$ is the gas adiabatic index.

{
Following \citet{kang2010}}, we parameterize the model as follows:
(1) The CR proton spectrum follows the test-particle DSA power-law, as $f_{\rm CR}(p)\propto p^{-q}$, where $q=3r/(r-1)$.
(2) The transition from the postshock thermal to CR spectra occurs at the injection momentum
\begin{equation}
p_{\rm inj}= Q_{\rm i} \cdot p_{\rm th,p},
\label{IP}
\end{equation}
where $Q_{\rm i}$ is the injection parameter.
{The main improvement here is that 
the postshock temperature, $T_2$, decreases slightly from $T_{2,0}$, and hence $p_{\rm th,p}$ does too, 
as the fraction of the shock energy transferred to CRs increases.}

Our model leads to the following form of the CR proton spectrum,
\begin{equation}
f_{\rm CR}(p) \approx \psi \cdot f_{\rm N} \left({p \over p_{\rm inj}}\right)^{-q} \exp \left[-\left({p \over p_{\rm max}}\right)^2\right].
\label{finj}
\end{equation}
Here, the maximum momentum of CR protons, $p_{\rm max}$, increases with the shock age \citep[e.g.,][]{kang2010}.
The normalization factor can be approximated as
\begin{equation}
f_{\rm N} = {n_2 \over \pi^{1.5}} p_{\rm th,p}^{-3} \exp(-Q_{\rm i}^2),
\label{fN}
\end{equation}
assuming the CR power-law spectrum is hinged to the postshock Maxwell distribution at $p_{\rm inj}$.
Therefore, in our model, $Q_{\rm i}$ is the key parameter that controls $f_{\rm N}$.
In addition, we introduce an additional parameter, $\psi\sim 1$, to accommodate any uncertainties in determining the value of $Q_{\rm i}$ and the resulting amplitude, $f_{\rm N}$.
Throughout this paper, however, $\psi= 1$ is used.
Figure \ref{f2}(a) shows the model spectrum, $f_{\rm CR}(p)$, calculated with Equations (\ref{finj})-(\ref{fN}), which illustrates the transition from the thermal to nonthermal CR spectra at $p_{\rm inj}$.

{
The PIC simulation described in section \ref{s2.2} indicates $Q_{\rm i} \approx 3$ when $p_{\rm max}/m_i c \approx 0.5$, but $Q_{\rm i}$ may further increase for $p_{\rm max}/m_i c \gg 1$, as noted above.
On the other hand, hybrid simulations for strong shocks of $\beta\approx 1$ (CS14a, CPS15) showed that it is expected to range as $Q_{\rm i} \approx 3.0-3.5$. 
As discussed in section \ref{s2.1}, higher $\beta$ could have two opposite effects on proton reflection, weaker magnetic mirror but stronger upstream turbulence.
So it is difficult to make quantitative predictions on the long-term evolution of $Q_i$ in high-$\beta$ shocks without performing plasma simulations of very long duration.
Here, we will consider the range of $Q_{\rm i}=3.3-3.5$ as an educated guess from the previous plasma simulation.
Moreover, in our analytic model, $Q_{\rm i} < 3.3$ would give the DSA efficiency of $\eta \gtrsim 0.01$ for $3\lesssim M_{\rm s} \lesssim 5$ (see Figure \ref{f4} below), 
which would be incompatible with the non-detection of $\gamma$-ray emission from galaxy clusters \citep{vazza2016}.}

\begin{figure*}[t]
\vskip -0.8 cm
\hskip -0.2 cm
\centerline{\includegraphics[width=1.05\textwidth]{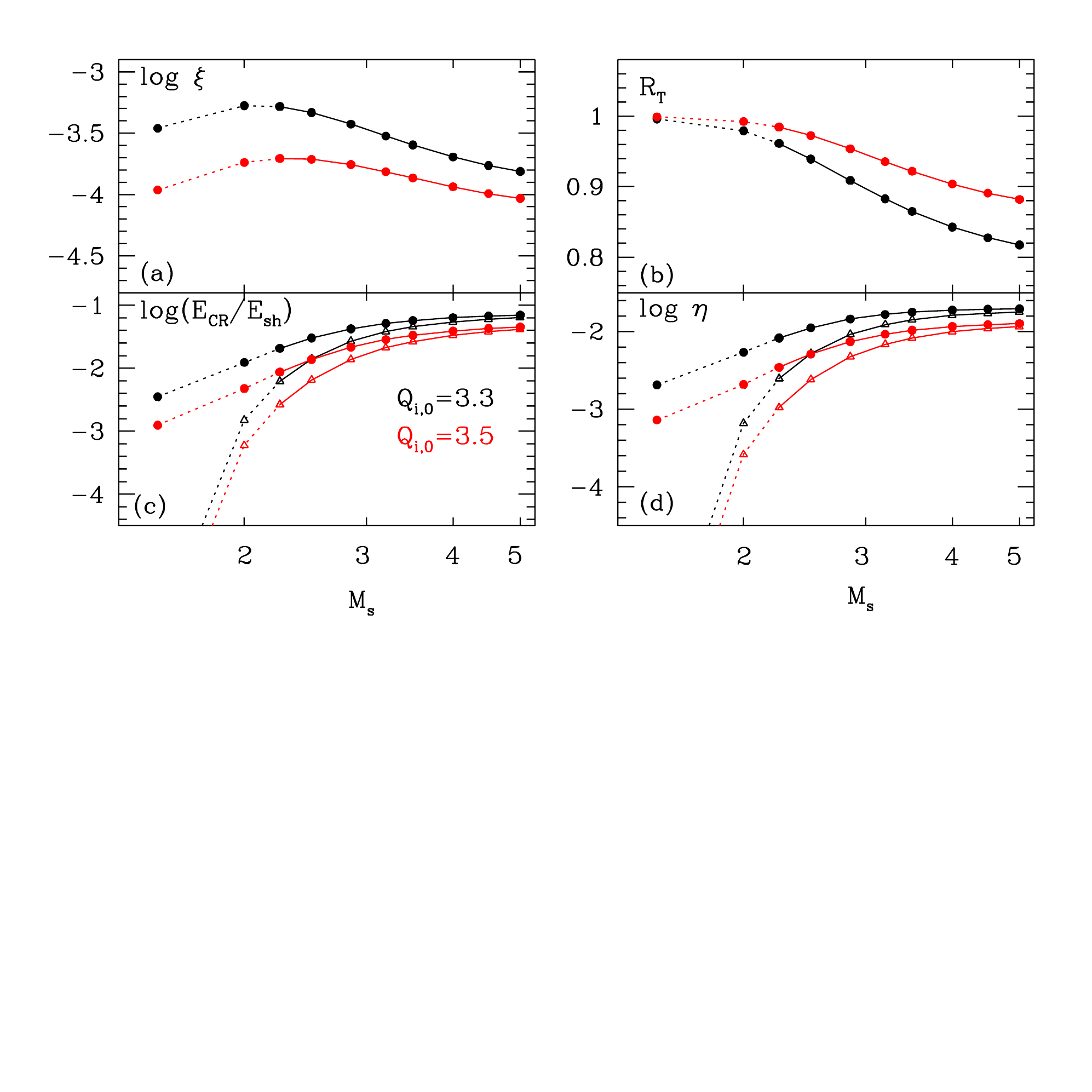}}
\vskip -8.4 cm
\caption{\label{f4}
The injection fraction, $\xi$, the temperature reduction factor, $R_{\rm T}$, the postshock CR energy fraction, 
$E_{\rm CR,2}/E_{\rm sh}$, and the CR acceleration efficiency, $\eta$, as a function of $M_{\rm s}$,
for $p_{\rm min}=p_{\rm inj}$ and $p_{\rm max}=10^5 m_p c$. Here, $T_1=10^8$ K.
The black and red filled circles connected with solid lines are the results for $Q_{\rm i,0}=3.3$ and $3.5$, respectively.
The two points for $M_{\rm s}=1.5$ and $2.0$ are connected with the dotted lines, 
because subcritical shocks with $M_{\rm s}<2.25$ may not preaccelerate
and inject thermal protons to the full DSA process according HRKM18.
The open triangles represent the values calculated with $p_{\rm min}=780~{\rm MeV}/c$.
}
\end{figure*}

From the model $f_{\rm CR}(p)$ in Equation (\ref{finj}), we calculate the injection fraction of CR protons by
\begin{equation}
\xi \equiv \frac{4\pi}{n_2}\int_{p_{\rm min}}^{p_{\rm max}} f_{\rm CR}(p) p^2 dp,
\label{inj}
\end{equation}
as in HRKM18.
The postshock CR energy density is estimated by
\begin{equation}
E_{\rm CR,2} = 4 \pi c \int_{p_{\rm min}}^{p_{\rm max}} (\sqrt{p^2+ (m_pc)^2}-m_pc) f_{\rm CR}(p) p^2 dp.
\label{ECR}
\end{equation}
In the case of very weak shocks, where the CR spectrum is dominated by low energy particles,
both $\xi$ and $E_{\rm CR,2}$ depend sensitively on the lower bound of the integrals, $p_{\rm min}$ \citep[e.g.,][]{pfrommer2004}.
We here adopt $p_{\rm min}\approx p_{\rm inj}$ for fiducial models, while $p_{\rm min}=780\ {\rm MeV}/c$, the
threshold energy of $\pi$-production reaction, will be considered as well for comparison.

As mentioned above, $E_{\rm CR,2}$ may increase, as $f_{\rm CR}(p)$ extends to higher $p_{\rm max}$, resulting in the decrease of the postshock gas temperature from $T_{2,0}$ to $T_2$
{(see Figure 5 of \citet{kang2005} and Figure 1 of CS14a).}
Thus, we introduce the temperature reduction factor,
\begin{equation}
R_{\rm T}= {{E_{\rm th}(T_2,0) - E_{\rm CR,2}}\over {E_{\rm th}(T_{2,0})} }.
\label{RN}
\end{equation}
Then, $T_2= R_{\rm T} T_{2,0}$ is the reduced postshock temperature.\footnote{The fraction of thermal particles that becomes CR protons is assumed to be small, i.e., $\xi \ll 1$.}

{CPS15 suggested that when the postshock CR energy density approaches to 
$E_{\rm CR,2} \approx 0.1 E_{\rm sh}=0.1 (\rho_1 u_{\rm sh}^2/2)$, the subshock weakens substantially, 
which suppresses the proton reflection and injection.
Hence, the normalization of $f_{\rm CR}$ is expected to decrease as $p_{\rm max}$ increases.}
Our model is designed to mimic such a behavior by finding the self-consistent postshock thermal distribution with a lower temperature, 
while $p_{\rm inj}$ is assumed to be fixed.
Then, the injection parameter increases as $Q_{\rm i} =  Q_{\rm i,0}/\sqrt{R_{\rm T}}$, where $Q_{\rm i,0}$ is the initial value, leading to smaller values of $f_{\rm N}$.
Note that $p_{\rm inj}$ at shocks with different parameters ($M_s$, $\theta_{\rm Bn}$, and $\beta$) is controlled by a number of complex kinetic process, and hence should be studied through long-term plasma simulations, beyond the current computational capacity.
Considering that the proton injection into DSA is yet to be fully understood, 
fixing $p_{\rm inj}$ while slightly increasing $Q_{\rm i}$ in our model should be regarded as a reasonable assumption.

Figure \ref{f2}(a) shows the model spectrum, including that of the self-consistent thermal distribution, in a $M_{\rm s}=3.2$ shock for $Q_{\rm i,0}=3.0-3.5$; 
the spectrum depends on the adopted value of $Q_{\rm i,0}$.
Panels (b)-(d) illustrate the change of the model spectrum as $p_{\rm max}$ increases in shocks with $M_{\rm s} = 2.5,$ 3.2, and 4.0, respectively.
As $p_{\rm max}$ and also $E_{\rm CR,2}$ increase, 
the Maxwellian part shifts to slightly lower $T_2$, and $R_{\rm T}$ decreases accordingly.
Because $p_{\rm inj}$ is assumed to be fixed, $Q_{\rm i}$ increases and thus the normalization factor $f_{\rm N}$ decreases in our model.

Figure \ref{f3} shows the change of $\xi$, $R_{\rm T}$, $E_{\rm CR,2}/E_{\rm sh}$, and  $\eta$, calculated with Equations (\ref{finj})-(\ref{RN}), as $p_{\rm max}$ increases for {$Q_{\rm i,0}=3.3$ (dashed lines) and 3.5 (solid lines) in shocks with $M_{\rm s}=2.25-4.0$.}
The CR acceleration efficiency is related to the postshock CR energy density, as $\eta = E_{\rm CR,2}/r E_{\rm sh}$.
Figure \ref{f3}(b) plots how $R_T$ decreases, as $p_{\rm max}$ increases.
{
The injection fraction, $\xi$, increases with increasing $p_{\rm max}$ during the early acceleration phase, but decreases for $p_{\rm max}/m_pc \gg 1$.
The latter behavior results from the gradual reduction of $f_{\rm CR}(p_{\rm inj})$, which is caused by the self-adjustment of the shock structure, that is, the
cooling of the postshock thermal protons, the growing of the precursor, and the weakening of the subshock due to the dynamical feedback of the CR pressure.}
$E_{\rm CR,2}/E_{\rm sh}$ and $\eta$, on the other hand, monotonically increase and approach to asymptotic values for $p_{\rm max}/m_pc \gtrsim 10^2$.

Figure \ref{f4} shows the asymptotic values of those quantities as a function of $M_{\rm s}$ (filled circles and lines)
for $Q_{\rm i,0}=3.3$ (black) and 3.5 (red), which would cover the most realistic range for ICM shocks (CS14a).
As mentioned in the introduction, HRKM18 showed that ICM $Q_{\parallel}$-shocks with
$M_{\rm s}<2.25$ may not inject protons into the DSA process, resulting in inefficient CR proton acceleration.
We here include the $M_{\rm s}= 1.5$ and $2.0$ cases (connected with dotted lines) for illustrative purposes, showing the values estimated with our model.

{Note that the asymptotic value of $\xi(M_{\rm s})$ decreases with increasing $M_{\rm s}$ for supercritical shocks with $M_{\rm s}\ge 2.25$.
This behavior is opposite to the relation, $\xi \propto M_{\rm s}^{1.5}$, during the very early acceleration stage of the PIC simulations reported in HRKM18.
In those PIC simulations, $p_{\rm max}/m_pc \lesssim 0.5$, so the CR feedback effect is not very significant.
However, our analytic model is designed to take account for the dynamic feedback of the CR pressure to the shock structure when $p_{\rm max}/m_pc \gg 1$,
so $\xi$ could be smaller at higher $M_{\rm s}$.

With the adopted value of $Q_{\rm i,0}=3.3-3.5$, $E_{\rm CR,2}/E_{\rm sh}< 0.1$, so the test-particle assumption should be valid.
The acceleration efficiency increases with $M_{\rm s}$ and is close to $\eta\approx0.01-0.02$ in the range of $M_{\rm s}=3-5$.
Obviously, if the injection parameter is larger than what the hybrid simulations of CS14a indicated, that is, $Q_{\rm i,0}>3.5$, then DSA would be even less efficient.}

In the studies of $\gamma$-ray emission from simulated galaxy clusters, the lower bound of $f_{\rm CR}$ 
is often taken as $p_{\rm min}=780~{\rm MeV}/c$, as noted above.
The open triangles in Figure \ref{f4} show $E_{\rm CR,2}/E_{\rm sh}$ and  $\eta$ calculated with this $p_{\rm min}$,
otherwise adopting the same analytic spectrum given in Equations (\ref{finj})-(\ref{fN}).
For $M_{\rm s}=2.25$, the acceleration efficiency with $p_{\rm min}=780~{\rm MeV}/c$ is smaller by a factor of 3.3 
than that with $p_{\rm min}=p_{\rm inj}$.
But the two estimations are similar for $M_{\rm s}\gtrsim 4$.
The efficiency with $p_{\rm min}=780~{\rm MeV}/c$ is $\eta\sim0.01$ in the range of $M_{\rm s}=3-5$, while $\eta\sim 10^{-3}$ for $M_{\rm s}=2.25$. 
If this result is extended to the case of $M_{\rm s} \sim 6$, $\eta$ would be still close to 0.01, which is about three times smaller than the efficiency reported by CS14a 
(i.e., $\eta \approx 0.036$ at $M_{\rm s} \sim 6.3$).
Note that this estimate is somewhat larger than the upper limit of $\eta\lesssim 10^{-3}$, quoted to be consistent with the non-detection of $\gamma$-ray emission from galaxy clusters by \citet{vazza2016}.
On the other hand, as HRKM18 shown, $\eta$ may be very small and negligible for shocks with $M_{\rm s}<2.25$, for which the fraction of the total shock dissipation in the ICM was shown to be substantial \citep[e.g.,][]{ryu2003}.
Hence, the consistency of our model for proton acceleration with the non-detection of cluster $\gamma$-rays
should be further examined
by considering the details of the characteristics of shocks in simulated galaxy clusters.

\section{Summary}
\label{s4}

{
The DSA efficiency for CR protons at low $M_{\rm s}$ $Q_\parallel$-shocks in the high-$\beta$ plasmas of the ICM has yet to be investigated through kinetic plasma simulations.
HRKM18 studied the injection and the early acceleration of protons up to $p_{\rm max}/m_ic \approx 0.5$ at such shocks through 1D PIC simulations, adopting reduced mass ratios of $m_i/m_e$. 
On the other hand, CS14a, CS14b, and CPS15 carried out hybrid simulations to study the DSA of protons, but considered only high $M_{\rm s}$ shocks in $\beta\approx 1$ plasmas.
Here, we revisited the test-particle DSA model for low $M_{\rm s}$ shocks with a thermal leakage injection recipe that was previously presented in \citet{kang2010}.
Reflecting new findings of recent plasma simulations, we improved the analytic DSA model by accounting for the
transfer of the postshock thermal energy to the CR energy and the weakening of the subshock due to the dynamical feedback of the CR pressure to the shock structure.}

We first set up an approximate analytic solution, $f_{\rm CR}(p)$, for CR protons in weak $Q_\parallel$-shocks.
We then calculated the injection fraction, $\xi$, the postshock CR energy fraction, $E_{\rm CR,2}/E_{\rm sh}$, and the acceleration efficiency, 
$\eta$, of CR protons.
The main aspects of our model and the main results are summarized as follows.

1. In weak shocks with $M_{\rm s} \lesssim 5$, above the injection momentum, $ p_{\rm inj}=Q_i\ p_{\rm th,p}$, $f_{\rm CR}(p)$ follows the test-particle DSA power-law, whose slope is determined by the shock compression ratio.

2. According to plasma simulations such as CS14a, CPS15, and HRKM18, as CR protons are accelerated to higher energies, the postshock gas temperature $T_2$ and the normalization of $f_{\rm CR}$ decreases (see Figure \ref{f2}).
Thus, in our model, while the injection momentum, $p_{\rm inj}$, is assumed to be fixed, the injection parameter increases as $Q_i= Q_{i,0}/\sqrt{R_{\rm T}}$, where $R_{\rm T}$ is the reduction factor of the postshock temperature. Then $Q_i$ determines the CR spectrum according to Equations (\ref{finj})-(\ref{RN}).
We adopt $Q_{i,0}\approx 3.3-3.5$, extrapolating the results of previous hybrid simulations.

{
3. In our model, as $f_{\rm CR}(p)$ extends to higher $p_{\rm max}/m_pc\gg 1$, $\xi$ first increases and then decreases due to the reduction of $T_2$ and the increase of $Q_{\rm i}$, although $\eta$ monotonically increases and approaches a time-asymptotic value. 
Such a behavior was previously seen in fluid DSA simulations \citep[e.g.,][]{kang2002}.

4. Both $\xi$ and $E_{\rm CR,2}/E_{\rm sh}$ depend on $Q_{i,0}$ and also the lower bound of the integrals, $p_{\rm min}$, especially in the case of very weak shocks (see Figure \ref{f4}).
For $p_{\rm min}\approx p_{\rm inj}$ and $Q_{i,0}=3.5$, the CR acceleration efficiency ranges as $\eta\approx 3.5\times 10^{-3} - 0.01$ for $2.25\lesssim M_{\rm s}\lesssim 5.0$.
If $p_{\rm min}\approx780~{\rm MeV}/c$ is adopted, it decreases to $\eta\approx 1.1 \times 10^{-3}-0.01$ for the same Mach number range.
If $Q_{i,0}=3.3$ is adopted, $\eta$ becomes larger by a factor of $1.5-2$, compared to the case with $Q_{i,0}=3.5$.}

5. In subcritical shocks with $M_{\rm s}<2.25$, protons may not be efficiently injected into DSA, so we expect that $\eta$ would be negligible at these very weak shocks (HRKM18).

{
In a parallel paper \citep{ha2019}, we will investigate the $\gamma$-ray emission as well as the neutrino emission from simulated galaxy clusters due to the inelastic collisions of CR protons and ICM thermal protons, 
based on the analytic CR proton spectrum proposed in this paper.
In particular, we will check whether the prediction for $\gamma$-ray emission complies with the upper limits imposed by Fermi LAT observations.}

\acknowledgments
{
We thank the anonymous referee for critical comments that help us improve this paper from its initial form.}
D.R. and J.-H. H. were supported by the National Research Foundation of Korea (NRF) through grants 2016R1A5A1013277 and 2017R1A2A1A05071429.
H.K. was supported by the Basic Science Research Program of the NRF through grant 2017R1D1A1A09000567.

\end{document}